\documentclass[11pt]{article}
\usepackage{graphicx}
\usepackage{epstopdf}
\usepackage[psamsfonts]{amssymb}
\usepackage{url}
\newtheorem{theorem}{Theorem}
\newtheorem{definition}{Definition}

\newtheorem{corollary}{Corollary}

\newcommand{\sps}{(\Sigma,{\cal L},\xi)}
\newcommand{\qed}{\mbox{} \hfill $\Box$}
\title{State Property Systems and Closure Spaces: Extracting the \\ Classical
and Nonclassical Parts\footnote{Published as: Aerts, D. and Deses, D., 2002, State property systems and closure spaces: extracting the classical and nonclassical parts, in {\it Probing the Structure of Quantum Mechanics: Nonlinearity, Nonlocality, Probability and Axiomatics}, eds. Aerts, D, Czachor, M. and Durt, T, World Scientific, Singapore.}}
\author{Diederik Aerts\\
        \normalsize\itshape
        Center Leo Apostel for Interdisciplinary Studies (CLEA) \\
         \normalsize\itshape
         Foundations of the Exact Sciences (FUND) \\
		\normalsize\itshape
 		Department of Mathematics \\
        \normalsize\itshape
        Vrije Universiteit Brussel, 1160 Brussels, 
       Belgium \\
        \normalsize
        E-Mail: \textsf{diraerts@vub.ac.be} \\ \\
		Didier Deses \\
		 \normalsize\itshape
		 Topology (TOPO) \\
		 \normalsize\itshape
		Foundations of the Exact Sciences (FUND) \\
		 \normalsize\itshape
		  Department of Mathematics\\
		   \normalsize\itshape
		   Vrije Universiteit Brussel, 1160 Brussels, 
       Belgium  \\
		   \normalsize
        E-Mail: \textsf{diddesen@vub.ac.be}
		}
\date{}
\begin{document}
\maketitle

\begin{abstract}
\noindent
In \cite{sp} an equivalence of the categories {\bf SP} and {\bf
Cls} was proven. The category {\bf SP} consists of the state
property systems \cite{aerts99} and their morphisms, which are the
mathematical structures that describe a physical entity by means
of its states and properties
\cite{aerts82,aerts83,aerts94,piron76,piron89,piron90}. The category {\bf
Cls} consists of the
closure spaces and the continuous maps.
In earlier work it has been shown, using the equivalence between {\bf
Cls} and {\bf SP}, that some of the axioms of quantum axiomatics
are equivalent with separation axioms on the corresponding closure
space. More particularly it was proven that the axiom of
atomicity is equivalent to the T$_{1}$ separation axiom \cite{ann}. In the
present article we analyze the intimate relation
that exists between classical and nonclassical in the state
property systems and disconnected and connected in the
corresponding closure space, elaborating results that appeared in
\cite{advv2001a,advv2001b}.
We introduce classical properties using the concept of super
selection rule, i.e. two properties are separated by a
superselection rule iff there do not exist `superposition states'
related to these two properties. Then we show that the classical
properties of a state property system correspond exactly to the
clopen subsets of the corresponding closure space. Thus connected
closure spaces correspond precisely to state property systems for
which the elements $0$ and $I$ are the only classical properties,
the so called pure nonclassical state property systems.
The main result is a decomposition theorem, which allows us to
split a state property system into a number of `pure nonclassical
state property systems' and a `totally classical state property
system'. This decomposition theorem for a state property system is the
translation of a decomposition theorem for the
corresponding closure space into its connected components.
\end{abstract}

\section{State Property Systems and Closure Spaces}
The general approaches to quantum mechanics make use of mathematical
structures that allow the description of pure quantum entities and pure
classical entities, as well as mixtures of both. In this article we study the
Geneva-Brussels approach, where the basic physical concepts are the one of
state and property of a physical entity
\cite{aerts82,aerts83,aerts94,piron76,piron89,piron90}. Traditionally the
collection
of properties is considered to be a complete lattice, partially ordered by the
implication of properties, with an orthocomplementation, representing the
quantum generalization of the `negation' of a property. A state is represented
by the collections of properties that are actual whenever the entity is in
this state. We mention however that in these earlier approaches
\cite{aerts82,aerts83,aerts94,piron76,piron89,piron90} the
mathematical structure that underlies the physical theory had not
completely been identified. To identify
the mathematical structure in a complete way, the structure of a state
property system was introduced in \cite{aerts99}.

Suppose that we consider a physical entity $S$, and we denote its set of
states by $\Sigma$ and its set of properties by ${\cal L}$. The state
property system
corresponding to this physical entity $S$ is a triple $(\Sigma,{\cal
L},\xi)$, where
$\Sigma$ is the set of states of $S$, ${\cal L}$ the set of properties of
$S$, and $\xi$ a
map from $\Sigma$ to ${\cal P}({\cal L})$, that makes correspond to each state
$p \in \Sigma$ the set of properties $\xi(p) \in {\cal P}({\cal L})$ that are
actual if the entity $S$ is in state $p$. Some additional requirements, that
express exactly how the physicists perceives a physical entity in relation with
its states and properties, are satisfied in a state property system. Let
us introduce the formal definition of a state property system and then explain
what these additional requirements mean.

\begin{definition}[State Property System]
\label{def:statprop} A triple $(\Sigma,{\cal L},\xi)$ is called a
state property system if $\Sigma$ is a set, ${\cal L}$ is a
complete lattice and $\xi :  \Sigma \rightarrow {\cal P}({\cal
L})$ is a function such that for $p \in \Sigma$, $0$ the minimal
element of ${\cal L}$ and $(a_i)_i \in {\cal L}$, we have:
\begin{eqnarray}
0 &\not\in& \xi(p) \label{001}\\
a_i \in \xi(p)\ \forall i &\Rightarrow&
\wedge_i a_i \in \xi(p) \label{eq:xi_inf} \label{002}
\end{eqnarray}
and for $a, b\in {\cal L}$ we have:
\begin{eqnarray}
a < b &\Leftrightarrow& \forall r \in \Sigma:a \in \xi(r) {\rm \
then}\ b \in \xi(r) \label{eq:xi2} \label{003}
\end{eqnarray}
\end{definition}
\noindent
We demand that ${\cal L}$, the set of properties, is a complete lattice.
This means that the set of properties is partially ordered, with the
physical meaning of
the partial order relation $<$ being the following: $a, b \in {\cal L}$, such
that $a < b$ means that whenever property $a$ is actual for the entity $S$,
also property $b$ is actual for the entity $S$. If ${\cal L}$ is a complete
lattice, it means that for an arbitrary family of properties $(a_i)_i \in {\cal
L}$ also the infimum $\wedge_ia_i$ of this family is a property. The property
$\wedge_ia_i$ is the property that is actual iff all of the properties $a_i$
are actual. Hence the infimum represents the logical `and'. The minimal element
$0$ of the lattice of properties is the property that is never actual ({\it
e.g.} the physical entity does not exist). Requirement (\ref{001})
expresses that a property that is in the image by $\xi$ of an arbitrary
state $p \in \Sigma$ can never be the $0$ property. Requirement (\ref{002})
expresses that if for a state $p \in \Sigma$ all the properties $a_i$ are
actual, this implies that for this state $p$ also the `and' property
$\wedge_ia_i$ is actual. Requirement (\ref{003}) expresses the meaning of the
partial order relation that we gave already: $a < b$ iff whenever $p$ is a
state of $S$ such that $a$ is actual if $S$ is in this state, then also
$b$ is actual if $S$ is in this state.

Along the same lines, just traducing what the physicist means when he
imagines the situation of two physical entities, of which one is a sub entity
of the other, the morphisms of state property systems can be deduced. More
concretely, suppose that $S$ is a sub entity of $S'$. Then each state $p'$ of
$S'$ determines a state $p$ of $S$, namely the state $p$ where the sub entity
$S$ is in when $S'$ is in state $p'$. This defines a map $m: \Sigma' \to
\Sigma$. On the other hand, each property $a$ of $S$ determines a property $a'$
of $S$, namely the property of the sub entity, but now conceived as a property
of the big entity. This defines a map $n: \Sigma \to \Sigma'$. Suppose that we
consider now a state $p'$ of $S'$, and a property $a$ of $S$, such that $a \in
\xi(m(p'))$. This means that the property $a$ is actual if the sub entity $S$
is in state $m(p')$. This state of affairs can be expressed equally by stating
that the property $n(a)$ is actual when the big entity is in state $p'$. Hence,
as a basic physical requirement of merological covariance we should have:
\begin{equation}
a \in \xi(m(p')) \Leftrightarrow n(a) \in \xi'(p')
\end{equation}
This all gives rise to the following definition of morphism for state
property systems.

\begin{definition}[Morphisms of State Property Systems]
Suppose that $(\Sigma,{\cal L},\xi)$ and $(\Sigma',{\cal L}',\xi')$ are
state property systems then
$$(m,n):(\Sigma',{\cal L}',\xi') \to (\Sigma,{\cal L},\xi)$$
is called an {\bf SP}-morphism if $m :
\Sigma' \to \Sigma$ and $n : {\cal L} \to {\cal L}'$ are functions
such that for $a \in {\cal L}$ and $p' \in \Sigma' :$
\begin{eqnarray}
a \in \xi(m(p')) \Leftrightarrow n(a) \in \xi'(p')
\end{eqnarray}
\end{definition}
\noindent
Using the previous definitions we can use these concept to generate a category
of state property systems, in the mathematical sense.

\begin{definition} [The Category SP]
The category of state property systems and their morphisms is
denoted by ${\bf SP}$.
\end{definition}

\begin{definition} [The Cartan Map]
If $(\Sigma,{\cal L},\xi)$ is a state
property system then its Cartan map is the mapping \mbox{$\kappa : {\cal
L} \to {\cal P}(\Sigma)$} defined by :
\begin{equation}
\kappa: {\cal L} \rightarrow {\cal P}(\Sigma):a \mapsto \kappa(a)
= \{p\in \Sigma \ \vert\ a \in \xi(p)\}
\end{equation}
\end{definition}
\noindent
It was amazing to be able to prove (see \cite{sp}) that this category of
states property systems and its morphisms is equivalent to a category which
arises as a generalization of the category of topological spaces and
continuous maps, namely to the category of closure spaces and the continuous
maps. We will now introduce this category of closure spaces.

A topological space consists of a set $X$, and a collection of `open' subsets,
such that $X$ is open, any union of open subsets is again open and any finite
intersection of open subsets is again open. A subset of $X$ is called
closed if it's complement is open. Therefore we have that in a topological
space the empty set is closed, any intersection of closed sets is closed and
any finite union of closed sets is again closed. Hence a topological space is
also defined by it's closed sets.

In mathematics the concept topological space is very useful and arises in many
different areas. However there are occasions when we `almost' have a
topological
space. Let's take the following example. Consider the plane $\mathbb{R}^2$ and
the collection of all convex subsets of $\mathbb{R}^2$ ($A$ is convex if the
segment between any two points of $A$ lies completely within $A$). Clearly
$\emptyset$ is convex and every intersection of convex sets is again convex.
However a finite union of convex sets does not need to be convex. Hence the
convex subsets of the plane can `almost' be considered as closed sets, but they
do not form a topological space. To be able to consider such structures one has
introduced the notion of closure spaces.

\begin{definition} [Closure Space]
\label{def:clos} A closure space $(X,{\cal F})$ consists of a set
$X$ and a family of subsets \mbox{${\cal F} \subseteq {\cal
P}(X)$} satisfying the following conditions:
\begin{eqnarray*}
\emptyset &\in& {\cal F} \\ (F_i)_i \in {\cal F} &\Rightarrow&
\cap_iF_i \in {\cal F}
\end{eqnarray*}
The closure operator corresponding to the closure space $(X,{\cal
F})$ is defined as
\begin{equation}
cl: {\cal P}(X) \rightarrow {\cal P}(X): A \mapsto \bigcap \{F \in
{\cal F}\ \vert\ A \subseteq F\}
\end{equation}
If $(X,{\cal F})$ and $(Y,{\cal G})$ are closure spaces then a
function $f : (X,{\cal F}) \to (Y,{\cal G})$ is called a
continuous map if $\forall B \in {\cal G}:f^{-1}(B) \in {\cal F}$.
The category of closure spaces and continuous maps is denoted by
${\bf Cls}$.
\end{definition}

\noindent
The following theorem shows how we can associate with each state
property system a closure space and with each morphism a continuous
map, hence we get the categorical equivalence described in
\cite{sp}.

\begin{theorem}
\label{corrF} The correspondence $F : {\bf SP} \longrightarrow
{\bf Cls}$ consisting of
\newline (1) the mapping
\begin{eqnarray*}
|{\bf SP}| &\rightarrow& |{\bf Cls}| \\ (\Sigma,{\cal L},\xi)
&\mapsto& F(\Sigma,{\cal L},\xi) = (\Sigma,\kappa({\cal L}))
\end{eqnarray*}
(2) for every pair of objects $(\Sigma,{\cal
L},\xi),(\Sigma',{\cal L}',\xi')$ of $\bf SP$ the mapping
\begin{eqnarray*}
{\bf SP}((\Sigma',{\cal L}',\xi'),(\Sigma,{\cal L},\xi))
&\rightarrow& {\bf Cls}(F(\Sigma',{\cal L}',\xi'),F(\Sigma,{\cal
L},\xi)) \\ (m,n) &\mapsto& m
\end{eqnarray*}
is a covariant functor.
\end{theorem}
\noindent
We can also connect a state property system to a closure space and
a morphism to a continuous map.

\begin{theorem} \label{corrG}
The correspondence $G : {\bf Cls} \longrightarrow {\bf SP}$
consisting of
\newline (1) the mapping
\begin{eqnarray*}
|{\bf Cls}| &\rightarrow& |{\bf SP}| \\ (\Sigma,{\cal F})
&\mapsto& G(\Sigma,{\cal F}) = (\Sigma,{\cal F},\bar \xi)
\end{eqnarray*}
where $\bar \xi : \Sigma \rightarrow {\cal P}({\cal F}) :  p \mapsto
\{F \in {\cal F}\ \vert\ p \in F\}$
\newline (2) for every pair of objects $(\Sigma,{\cal F}),(\Sigma',
{\cal F}')$ of $\bf Cls$ the mapping
\begin{eqnarray*}
{\bf Cls}((\Sigma',{\cal F}'),(\Sigma,{\cal F})) &\rightarrow&
{\bf SP}(G(\Sigma',{\cal F}'),G(\Sigma,{\cal F})) \\ m &\mapsto&
(m,m^{-1})
\end{eqnarray*}
is a covariant functor.
\end{theorem}

\begin{theorem}[Equivalence of $\bf SP$ and $\bf Cls$]
\label{theor:equiv} The functors
$$F:{\bf SP}\rightarrow {\bf Cls}$$
$$G:{\bf Cls} \rightarrow {\bf SP}$$
establish an equivalence of categories.
\end{theorem}
\noindent
The above equivalence is a very powerful tool for studying state property
systems. It states that the lattice $\mathcal{L}$ of properties can be seen as
the lattice of closed sets of a closure space on the states $\Sigma$,
conversely every closure space on $X$ can be considered as a set of states
($X$) and a lattice of properties (the lattice of closed sets).

Recall that closure spaces are in fact a generalization of topological spaces,
hence a number of topological properties have been generalized to closure
spaces. Moreover with the previous equivalence, a concept which can be defined
using closed sets on a closure space can be translated in an equivalent concept
for state property systems. At first sight this translation does not need to be
meaningful in the context of physical systems. However it turned out that many
such translations actually coincided with well known physical concepts.

We shall give one example which was studied in \cite{ann}.
A topological space is called $T_1$ if the following separation axiom is
satisfied. For every two points $x,y$ there are open sets which contain $x$
resp. $y$ but do not contain $y$ resp. $x$. This is equivalent to stating that
all singletons are closed sets. Hence the following definition.

\begin{definition}[$T_1$ Closure Space]
A closure space $(X,\mathcal{F})$ is a $T_1$ closure space iff
$\forall x\in X:\{x\}\in \mathcal{F}$.
\end{definition}
\noindent
In the theory of state property systems, or more general of property lattices
the concept of atomistic lattice is quite fundamental. In \cite{ann} it was
proven that using the equivalence between state property systems and closure
spaces both concepts are in fact related.

\begin{definition}[Atomistic State Property System]
Let $(\Sigma,{\cal L},\xi)$ be a state property system. Then the
map $s_{\xi}$ maps a state $p$ to the strongest property it makes
actual, i.e.
\begin{equation}
s_{\xi} : \Sigma \to {\cal L} : p \mapsto \wedge \xi(p)
\end{equation}
T.F.A.E.
\begin{itemize}
\item[(1)]{$\xi:\Sigma \to \mathcal{P}(\mathcal{L})$ is injective and
$\forall p\in \Sigma:s_\xi(p)$ is an atom of $\mathcal{L}$.}
\item[(2)]{$\forall p,q\in \Sigma:\xi(p)\subset \xi(q)\Rightarrow
p=q$}
\item[(3)]{$F(\Sigma,\mathcal{L},\xi)=(\Sigma,\kappa(\mathcal{L}))$ is
a $T_1$ closure space.}
\end{itemize}
If a state property system satisfies one, and hence all of the
above conditions it is called an atomistic state property system,
in this case $\mathcal{L}$ is a complete atomistic lattice.
\end{definition}
\noindent
If we write $\bf Cls_1$ for the full subcategory of $\bf Cls$ given by $T_1$
closure spaces, and $\bf SP_a$ for the full subcategory of $\bf SP$ given
by the
atomistic state property systems, then the general equivalence can be reduced.

\begin{theorem}[Equivalence of $\bf SP_a$ and $\bf Cls_1$]
The functors
$$F:{\bf SP_a}\rightarrow {\bf Cls_1}$$
$$G:{\bf Cls_1} \rightarrow {\bf SP_a}$$
establish an equivalence of categories.
\end{theorem}
\noindent
For a more extensive study of separation axioms and their relation with state
property systems we refer to \cite{annths}. In the present text our final aim
is to use the described equivalence to translate the concept of connectedness
in closure spaces into terms of state property systems. It will give us a means
to distinguish `classical' and `quantum mechanical' properties of a physical
entity. First we will need a more precise concept of classical property.

\section{Super Selection Rules}

In this section we start to distinguish the classical aspects of the
structure from the quantum aspects. We all know that the concept of
superposition state is very important in quantum mechanics. The superposition
states are the states that do not exist in classical physics and hence their
appearance is one of the important quantum aspects. To be able to define
properly a superposition state we need the linearity of the set of states. On
the level of generality that we work now, we do not necessarily have this
linearity, which could indicate that the concept of superposition state cannot
be given a meaning on this level of generality. This is however not really
true: the concept can be traced back within this general setting, by
introducing the idea of `superselection rule'. Two properties are separated by
a superselection rule iff there do not exist `superposition states' related to
these two properties. This concept will be the first step towards
a characterization of classical properties of a physical system.

\begin{definition} [Super Selection Rule]
Consider a state property system $(\Sigma,{\cal L},\xi)$. For $a,
b \in {\cal L}$ we say that $a$ and $b$ are separated by a super
selection rule, and denote $a$ ssr $b$, iff for $p \in \Sigma$ we
have:
\begin{equation}
a \vee b \in \xi(p) \Rightarrow a \in \xi(p) {\rm \ or}\ b \in
\xi(p)
\end{equation}
\end{definition}
\noindent
We again use the equivalence between state property systems and closure
spaces to translate the concept of `separation by a superposition rule' into a
concept for the closed sets of a closure space. Amazingly we find that
properties that are `separated by a superselection rule' (i.e. they are
`classical' properties in a certain sense) correspond to closed sets that also
behave in a classical way, where classical now refers to classical topology.

\begin{theorem}
Consider a state property system $(\Sigma,{\cal L},\xi)$ and its
corresponding closure space ${\cal F} = \kappa({\cal L})$. For $a
, b \in {\cal L}$ we have:
\begin{equation}
a \ ssr\  b \Leftrightarrow \kappa(a \vee b) = \kappa(a) \cup
\kappa(b) \Leftrightarrow \kappa(a) \cup \kappa(b) \in {\cal F}
\end{equation}
\end{theorem}

\noindent
Proof: Suppose that $a, b \in {\cal L}$ such that $a\ ssr\ b$. If $p \in
\kappa(a \vee b)$, then $a \vee b \in \xi(p)$. Then it follows
that $a \in \xi(p)$ or $b \in \xi(p)$. So we have $p \in
\kappa(a)$ or $p \in \kappa(b)$, which shows that $p \in \kappa(a)
\cup \kappa(b)$. This proves that $\kappa(a \vee b) \subseteq
\kappa(a) \cup \kappa(b)$. We obviously have the other inclusion
and hence $\kappa(a \vee b) = \kappa(a) \cup \kappa(b)$. It
follows immediately that $\kappa(a) \cup \kappa(b) \in {\cal F}$.
Conversely, if $\kappa(a) \cup \kappa(b) \in {\cal F}$, then there
exists a property $c \in {\cal L}$ such that $\kappa(c) =
\kappa(a) \cup \kappa(b)$. From $\kappa(a) \subseteq \kappa(c)$ it
follows that $a < c$, and in a similar way we have $b < c$. So it
follows that $a \vee b < c$. As a consequence we have $\kappa(a
\vee b) \subseteq \kappa(c) = \kappa(a) \cup \kappa(b)$. Since
$\kappa(a) \cup \kappa(b) \subseteq \kappa(a \vee b)$, we have
$\kappa(a \vee b) = \kappa(a) \cup \kappa(b)$. Consider now an
arbitrary $p \in \Sigma$ such that $a \vee b \in \xi(p)$. Then $p
\in \kappa(a \vee b) = \kappa(a) \cup \kappa(b)$. As a consequence
$p \in \kappa(a)$ or $p \in \kappa(b)$. This proves that $a \in
\xi(p)$ or $b \in \xi(p)$ which shows that $a$ ssr $b$.
\qed

\medskip
\noindent
This theorem shows that the properties that are separated by a
super selection rule are exactly the ones that behave also
classically within the closure system. In the sense that their set
theoretical unions are closed. This also means that if our closure
system reduces to a topology, and hence all finite unions of
closed subsets are closed, all finite sets of properties are
separated by super selection rules.

\begin{corollary}
Let $\sps$ be a state property system. T.F.A.E.:
\begin{itemize}
\item[(1)] Every two properties of $\mathcal{L}$ are separated by a super
selection
rule.
\item[(2)] The corresponding closure space
$(\Sigma,\kappa(\mathcal{L}))$ is a topological space.
\end{itemize}
A state property satisfying one, and hence both of the above conditions will be
called a `super selection classical' state property system or `s-classical'
state property system. The full
subcategory of $\bf SP$ given by the s-classical state
property systems will be written as $\bf ^{sc} SP$.\end{corollary}

\noindent
Hence the equivalence between state property systems and closure space can be
reduced to an equivalence between s-classical state property systems, in
which no two properties have `superposition states' related to them, and
topological spaces.

\begin{theorem}[Equivalence of $\bf ^{sc} SP$ and $\bf Top$]
The functors
$$F:{\bf ^{sc} SP}\rightarrow {\bf Top}$$
$$G:{\bf Top} \rightarrow {\bf ^{sc} SP}$$
establish an equivalence of categories.
\end{theorem}

\section{D-classical Properties}

We are ready now to introduce the concept of a `deterministic classical
property' or `d-classical property'. To make clear what we mean by this we
have to explain shortly how properties are tested.
For each property $a \in {\cal L}$ there exists a test $\alpha$, which is
an experiment that can be performed on the physical
entity under study, and that can give two outcomes, `yes' and `no'. The
property $a$ tested by the experiment $\alpha$ is actual
iff the state $p$ of $S$ is such that we can predict with certainty
(probability equal to 1) that the outcome `yes' will occur
for the test
$\alpha$. If the state $p$ of $S$ is such that we can predict with
certainty that the outcome `no' will occur, we
test in some way a complementary property of the property $a$, let us
denote the complementary property by $a^c$. Now we have
three possibilities: (1) the state of $S$ is such that $\alpha$ gives `yes'
with certainty; (2) the state of $S$ is
such that $\alpha$ gives `no' with certainty; and (3) the state of $S$ is
such that neither the outcome `yes' nor
the outcome `no' is certain for the experiment $\alpha$. The third case
represents the situations of `quantum indeterminism'.
That is the reason that a property $a$ tested by an experiment $\alpha$
where the third case is absent will be called a
`deterministic classical' property or `d-classical' property.

\begin{definition} [D-classical Property]
Consider a state property system $(\Sigma,{\cal L},\xi)$. We say
that a property $a \in {\cal L}$ is a `deterministic classical property' or
`d-classical' property, if
there exists a property $a^c \in {\cal L}$ such that $a \vee a^c =
I$, $a \wedge a^c = 0$ and $a$ ssr $a^c$.
\end{definition}
\noindent
Remark that for every state property system
$(\Sigma,{\cal L},\xi)$ the properties $0$ and $I$ are d-classical
properties. Note also that if $a \in {\cal L}$ is a d-classical
property, we have for $p \in \Sigma$ that $a \in \xi(p)
\Leftrightarrow a^c \notin \xi(p)$ and $a \notin \xi(p)
\Leftrightarrow a^c \in \xi(p)$. This follows immediately from the
definition of a d-classical property.

\begin{theorem}
Consider a state property system $(\Sigma,{\cal L},\xi)$. If $a
\in {\cal L}$ is a d-classical property, then $a^c$ is unique and is
a d-classical property. We will call it the complement of $a$.
Further we have:
\begin{eqnarray*}
(a^c)^c &=& a    \label{idempot} \\ a < b &\Rightarrow& b^c < a^c
\\ \kappa(a^c) &=& \kappa(a)^C
\end{eqnarray*}
\end{theorem}

\noindent
Proof: Suppose that we have another property $b \in {\cal L}$ such that
$a \vee b = I$, $a \wedge b = 0$ and $a$ ssr $b$. Consider an
arbitrary state $p \in \Sigma$ such that $a^c \in \xi(p)$. This
means that $a \notin \xi(p)$. We have however $a \vee b \in
\xi(p)$, which implies, since $a$ ssr $b$, that $a \in \xi(p)$ or
$b \in \xi(p)$. As a consequence we have $b \in \xi(p)$. This
means that we have proven that $a^c < b$. In a completely
analogous way we can show that also $b < a^c$, which shows that
$a^c$ is unique. Obviously $a^c$ is a d-classical property. Then the
idempotency follows from the fact that $a$ is the complement of
$a^c$ and from the uniqueness of the complement. Consider $a < b$
and an arbitrary state $p \in \Sigma$ such that $b^c \in \xi(p)$.
This means that $b \notin \xi(p)$, which implies that $a \notin
\xi(p)$. As a consequence we have $a^c \in \xi(p)$. So we have
shown that $b^c < a^c$. Further we have $p \in \kappa(a^c)$ iff
$a^c \in \xi(p)$. From the above mentioned remark this is
equivalent with $a \notin \xi(p)$. and $p \notin \kappa(a)$ which
is the same as saying that $p \in \kappa(a)^C$. So we have
$\kappa(a^c) = \kappa(a)^C$.
\qed

\begin{definition} [Connected Closure Space]
A closure space $(X,{\cal F})$ is called connected if the only
clopen (i.e. closed and open) sets are $\emptyset$ and $X$.
\end{definition}
\noindent
We shall see now that these subsets that make closure systems
disconnected are exactly the subsets corresponding to d-classical
properties.

\begin{theorem} \label{prop clasclop} Consider a state property system
$(\Sigma,{\cal L},\xi)$ and its corresponding closure space
$(\Sigma,\kappa({\cal L}))$. For $a \in {\cal L}$ we have:
\begin{equation}
a \ {\rm is\ d-classical}\ \Leftrightarrow \kappa(a)\ {\rm is\
clopen}
\end{equation}
\end{theorem}

\noindent
Proof:
>From the previous propositions it follows that if $a$ is
d-classical, then $\kappa(a)$ is clopen. So now consider a clopen
subset $\kappa(a)$ of $\Sigma$. This means that $\kappa(a)^C$ is
closed, and hence that there exists a property $b \in {\cal L}$
such that $\kappa(b) = \kappa(a)^C$. We clearly have $a \wedge b =
0$ since there exists no state $p \in \Sigma$ such that $p \in
\kappa(a)$ and $p \in \kappa(b)$. Since $\Sigma = \kappa(a) \cup
\kappa(b)$ we have $a \vee b = I$. Further we have that for an
arbitrary state $p \in \Sigma$ we have $a \in \xi(p)$ or $b \in
\xi(p)$ which shows that $a$ ssr $b$. This proves that $b = a^c$
and that $a$ is d-classical.
\qed

\medskip
\noindent
This means that the d-classical properties correspond
exactly to the clopen subsets of the closure system.

\begin{corollary}
\label{cor quantum} Let $\sps$ be a state property system. T.F.A.E.
\\ (1) The properties $0$ and $I$ are the only d-classical ones. \\
(2) $F\sps = (\Sigma, \kappa({\cal L}))$ is a connected closure
space.
\end{corollary}

\noindent
We now introduce `completely quantum mechanical' or pure nonclassical state
property systems,
in the sense that there are no (non-trivial) d-classical properties.

\begin{definition} [Pure Nonclassical State Property System]
A state property system $\sps$ is called a pure nonclassical state
property system if the properties $0$ and $I$ are the only
d-classical properties.
\end{definition}

\begin{theorem}
Let $(\Sigma,{\cal F})$ be a closure space. T.F.A.E. \\ (1)
$(\Sigma,\cal F)$ is a connected closure space. \\ (2)
$G(\Sigma,{\cal F}) = (\Sigma,{\cal F},\bar \xi)$ is a pure
nonclassical state property system.
\end{theorem}

\noindent
Proof:
Let $(\Sigma,{\cal F})$ be a connected state property system. Then
$\emptyset$ and $\Sigma$ are the only clopen sets in
$(\Sigma,{\cal F})$. Since the Cartan map associated to $\xi$ is
given by $\kappa : {\cal F} \to {\cal P}(\Sigma) : F \mapsto F$,
we have $\kappa(\emptyset) = \emptyset$ and $\kappa(\Sigma) =
\Sigma$. Applying proposition~\ref{prop clasclop}, we find that
$\emptyset$ and $\Sigma$ are the only d-classical properties of
${\cal F}$. Conversely, let $G(\Sigma,{\cal F}) = (\Sigma,{\cal
F},\bar \xi)$ be a pure nonclassical state property system. Then by
corollary~\ref{cor quantum}, $(\Sigma,{\cal F}) = FG(\Sigma,{\cal
F})$ is a connected closure space.
\qed

\medskip
\noindent
If we define $\bf SP_Q$ as the full subcategory of
{\bf SP} where the objects are the pure nonclassical state
property systems and we define $\bf Cls_{C}$ as the full
subcategory of {\bf Cls} where the objects are the connected
closure spaces, then the previous propositions and
theorem~\ref{theor:equiv} imply an equivalence of the categories
$\bf SP_{Q}$ and $\bf Cls_{C}$.

\begin{theorem}[Equivalence of $\bf SP_Q$ and $\bf Cls_C$]
The functors
$$F:{\bf SP_Q}\rightarrow {\bf Cls_C}$$
$$G:{\bf Cls_C} \rightarrow {\bf SP_Q}$$
establish an equivalence of categories.
\end{theorem}

\noindent
Again we have found using the equivalence \ref{theor:equiv} that a physical
concept (i.e. nonclassicality) translates to a known topological property
(i.e. connectedness). In the next section we will use topological methods to
construct a decomposition of a state property system into pure nonclassical
components.

\section{Decomposition Theorem}

As for topological spaces, every closure space can be decomposed
uniquely into connected components. In the following we say that,
for a closure space $(X,{\cal F})$, a subset $A \subseteq X$ is
connected if the induced subspace is connected. It can be shown
that the union of any family of connected subsets having at least
one point in common is also connected. So the component of an
element $x \in X$ defined by
\begin{equation}
K_{\bf Cls}(x)=\bigcup \{A\subseteq X \ | \ x \in A, A \hbox{
connected }\}
\end{equation}
is connected and therefore called the connection component of $x$.
Moreover, it is a maximal connected set in $X$ in the
sense that there is no connected subset of $X$ which properly
contains $K_{\bf Cls}(x)$. From this it follows that for closure
spaces $(X,{\cal F})$ the set of all distinct connection
components in $X$ form a partition of $X$. So we can consider the following
equivalence relation on $X$ : for $x,y \in X$ we say that $x$ is equivalent
with $y$ iff the connection components $K_{\bf Cls}(x)$ and $K_{\bf
Cls}(y)$ are equal. Further we remark that the connection components are
closed sets.

In the following we will try to decompose state property systems similarly into
different components.

\begin{theorem}
Let $\sps$ be a state property system and let $(\Sigma,\kappa({\cal L}))$
be the corresponding closure space.
Consider the following equivalence relation on $\Sigma$ :
\begin{equation}
p \sim q \Leftrightarrow K_{\bf Cls}(p)=K_{\bf Cls}(q)
\end{equation}
with equivalence classes $\Omega=\{\omega(p)|p\in \Sigma\}$.
If
$\omega \in \Omega$ we define the following :
\begin{eqnarray*}
\Sigma_\omega &=&\omega  = \{p \in \Sigma \ | \ \omega(p) = \omega\} \\
s(\omega)&=&s(\omega(p))=a, \textrm{ such
that } \kappa(a)=\omega(p) \\ {\cal
L}_\omega&=&[0,s(\omega)]=\{a\in {\cal L} \ | \ 0 \leq a \leq s(\omega)
\}\subset {\cal L}\\ \xi_\omega &:&\Sigma_\omega\to {\cal P}({\cal
L}_\omega):p \mapsto \xi(p)\cap {\cal L}_\omega
\end{eqnarray*}
then $(\Sigma_\omega,{\cal L}_\omega,\xi_\omega)$ is a state property
system.
\end{theorem}

\noindent
Proof:
Since ${\cal L}_\omega$ is a sublattice (segment) of ${\cal L}$,
it is a complete lattice with maximal element $I_\omega=s(\omega)$
and minimal element $0_\omega=0$. Let $p \in \Sigma_\omega$. Then
$0 \not \in \xi(p)$. So $0 \not \in \xi(p)\cap {\cal
L}_\omega=\xi_\omega(p)$. If $a_i \in \xi_\omega(p), \ \forall i$,
then $a_i \in {\cal L}_\omega$ and $a_i\in \xi(p), \ \forall i$.
Hence $\wedge a_i \in {\cal L}_\omega \cap \xi(p) =
\xi_\omega(p)$. Finally, let $a,b \in {\cal L}_\omega$ with
$a<_\omega b$ and let $r\in \Sigma_\omega$. If $a \in
\xi_\omega(r)$, then $a\in {\cal L}_\omega$ and $a\in \xi(r)$,
thus $b\in {\cal L}_\omega$ and $b\in \xi(r)$. So $b\in
\xi_\omega(r)$. Conversely, if $a,b \in {\cal L}_\omega$ and
$\forall r\in \Sigma_\omega: a\in \xi_\omega(r) \Rightarrow b\in
\xi_\omega(r)$ then we consider a $q$ such that $a\in \xi(q)$ ($q$
must be in $\Sigma_\omega$ by definition of ${\cal L}_\omega$).
Then $a\in \xi_\omega(q)$ implies that $b\in \xi_\omega(q)$. So
$b\in \xi(q)$ and $a < b$. Thus $a<_\omega b$.
\qed

\medskip
\noindent
Moreover we can show that the above introduced state property
systems $(\Sigma_{\omega},{\cal L}_{\omega},\xi_{\omega})$  have
no proper d-classical properties, and hence are pure nonclassical
state property systems.

\begin{theorem}
Let $(\Sigma,{\cal L},\xi)$ be a state property system. If $\omega
\in \Omega$, then $(\Sigma_\omega,{\cal L}_\omega,\xi_\omega)$ is
a pure nonclassical state property system.
\end{theorem}

\noindent
Proof:
If $a$ is classical element of ${\cal L}_\omega$, then $\kappa(a)$
must be a clopen set of the associated closure space
$(\Sigma_{\omega},\kappa({\cal L}_\omega))$ which
is a connected subspace of $(\Sigma,\kappa({\cal L}))$. Hence
there are no proper classical elements of ${\cal L}_\omega$.
\qed

\begin{theorem}Let $\sps$ be a state property system. If we introduce the
following :
\begin{eqnarray*}
\Omega&=&\{\omega(p) \ | \ p\in \Sigma\}\\ \mathcal{C}&=&\{\vee
s(\omega_i) \ | \ \omega_i\in \Omega\} \\ \eta&:&\Omega \to
{\cal P}(\mathcal{C}):\omega=\omega(p)\mapsto \xi(p)\cap \mathcal{C}
\end{eqnarray*}
then $(\Omega,{\cal C},\xi)$ is an atomistic state property system.
\end{theorem}

\noindent
Proof:
First we remark that $\eta$ is well defined because if
$\omega(p)=\omega(q)$, then $\xi(p) \cap {\cal C} = \xi(q) \cap {\cal C}$.
Indeed, if $\vee s(\omega_i)\in \xi(p)$ then $p \in \kappa(\vee s(\omega_i))
= cl( \cup\omega_i)$ in the corresponding closure space
$(\Sigma,\kappa({\cal L}))$. Since $cl(\cup\omega_i)$ is not connected we
have that $K_{\bf
Cls}(p)=\omega(p)=\omega(q)\subset cl(\cup\omega_i)$ so $q \in cl(
\cup\omega_i) = \kappa(\vee s(\omega_i))$ and $\vee s(\omega_i)\in \xi(q)$.
Now, since $\mathcal{C}$ is a sublattice of ${\cal L}$ it is a complete
lattice with $1_\mathcal{C}=1$ and $0_\mathcal{C}=0$. By definition
$\mathcal{C}$ is generated by its atoms $\{s(\omega) \ | \ \omega\in
\Omega\}$. Clearly $0\not \in \eta(\omega(p))$ because $0 \not \in
\xi(p)$. If $a_i\in \eta(\omega(p))=\xi(p) \cap
\mathcal{C}, \ \forall i$, then  $\wedge a_i\in \xi(p)\cap
\mathcal{C}=\eta(\omega(p))$. Finally, let $a,b \in {\cal C}$  with
$a<_\mathcal{C}
b$. Let $\omega(p) \in \Omega$ with $a \in \eta(\omega(p))$. Thus  $a \in
\xi(p)$. $a <_{\cal C} b$ implies $a < b$. So we have $b \in \xi(p) \cap
{\cal C} = \eta(\omega(p))$. Conversely, let $a,b\in \mathcal{C}$ and
assume that $\forall p\in \Sigma :
a\in\eta(\omega(p))\Rightarrow b \in\eta(\omega(p))$. Then we have
$\forall p\in \Sigma: a\in\xi(p)\Rightarrow b \in\xi(p)$. Thus $a<
b$ and $a<_\mathcal{C}b$. In order to prove that $(\Omega,{\cal C},\eta)$
is an atomistic state property, we show that $\eta$ is injective. So
consider $p, q \in \Sigma$ such that $\omega(p) \neq \omega(q)$. Since $p
\in \omega(p) = \kappa(s(\omega(p)))$, we have $s(\omega(p)) \in \xi(p)
\cap {\cal C}$ and since $q \notin \omega(p) = \kappa(s(\omega(p)))$ we
have $s(\omega(p)) \notin \xi(q) \cap {\cal C}$. This implies that $\xi(p)
\cap {\cal C} \neq \xi(q) \cap {\cal C}$, i.e. $\eta(\omega(p)) \neq
\eta(\omega(q))$. Thus $\eta$ is injective and $(\Omega,{\cal C},\eta)$ is
an atomistic state property system.
\qed

\begin{theorem}
$(\Omega,\mathcal{C},\eta)$ is a totally classical state property
system, in the sense that the only pure nonclassical segments
(i.e. segments with no proper classical elements) are trivial,
i.e. $\{0,s(\omega)\}$.
\end{theorem}

\noindent
Proof:
Suppose $[0,a]$ is a pure nonclassical segment of $\mathcal{C}$,
then in the corresponding closure space $(\Sigma,\kappa({\mathcal
L}))$ the subset $\kappa(a)$ is connected hence
$\kappa(a)\subset\omega$ for some $\omega\in \Omega$, hence
$a<s(\omega)$. Since $s(\omega)$ is an atom, $a = s(\omega)$. Thus
$[0,a] = \{0,s(\omega)\}$.
\qed

\begin{corollary}
The closure space associated with $(\Omega,\mathcal{C},\eta)$ is a
totally disconnected closure space.
\end{corollary}

\noindent
Summarizing the previous results we get:

\medskip
\begin{theorem}[decomposition theorem]
Any state property system $\sps$ can be decomposed into:
\begin{itemize}
\item a number of pure nonclassical state property systems
$(\Sigma_\omega,{\cal L}_\omega,\xi_\omega),\omega \in \Omega$
\item and a totally classical state property system $(\Omega,\mathcal{C},\eta)$
\end{itemize}
\end{theorem}

\noindent
Thus the decomposition of a closure space into its maximal connected components
yields a way to decompose a state property system $\sps$ into pure
nonclassical state property systems $(\Sigma_\omega,{\cal
L}_\omega,\xi_\omega),\omega \in \Omega$. In the context of closure spaces the
maximal connected components are subspaces of the given space. However we do
not yet have that the pure nonclassical state property systems
$(\Sigma_\omega,{\cal
L}_\omega,\xi_\omega)$ are subsystems of $\sps$. To show this we introduce a
new concept of subsystem.

\section{Closed Subspaces and $ap$-Subsystems}

\begin{definition} [AP-subsystem]
Let $(\Sigma,\mathcal{L},\xi)$ be a state property system and let
$a\in \mathcal{L}$. Consider the following:
\begin{itemize}
\item $\Sigma'=\kappa(a)$
\item $\mathcal{L}'=[0,a]$
\item $\xi'=\xi_{|\Sigma'}$
\end{itemize}
We now have a new state property system
$(\Sigma',\mathcal{L}',\xi')$ which we shall call an 'actual
property' ($ap$-) subsystem of $(\Sigma,\mathcal{L},\xi)$
generated by $a$ .
\end{definition}

\noindent
The name `actual property' subsystem comes from the physical
interpretation of this construction: give a property $a$ of the
physical system, we consider only those states $\Sigma'$ for which
$a$ is always actual.

\begin{theorem}
Let $(\Sigma',\mathcal{L}',\xi')$ be an $ap$-subsystem of
$(\Sigma,\mathcal{L},\xi)$, generated by $a$. Consider the
corresponding closure spaces $(\Sigma',\kappa(\mathcal{L}'))$ and
$(\Sigma,\kappa(\mathcal{L}))$, we have that
$(\Sigma',\kappa(\mathcal{L}'))$ is a closed subspace of
$(\Sigma,\kappa(\mathcal{L}))$.
\end{theorem}

\noindent
Proof:
Follows immediately from the definition.
\qed

\begin{theorem}
Consider a closed subspace $(\Sigma',\mathcal{F}')$ of the closure
space $(\Sigma,\mathcal{F})$, we have that
$(\Sigma',\mathcal{F}',\bar \xi')$ is an $ap$-subsystem of
$(\Sigma,\mathcal{F},\bar \xi)$ generated by $\Sigma'$.
\end{theorem}

\noindent
Proof:
Follows immediately from the definition.
\qed

\medskip
\noindent
>From the above two theorems we see that $ap$-subsystems correspond
exactly to closed subspaces of the associated closure space.

Any closed subspace $\Sigma'$ of a closure space
$(\Sigma,\mathcal{F})$ induces in a natural way a canonical
inclusion map:
$$i:(\Sigma',\mathcal{F}')\to (\Sigma,\mathcal{F})$$
which in turn, by the functional equivalence between the category
of closure spaces and state property systems gives a morphism:
$$ (i,i^{-1}):(\Sigma',\mathcal{F}',\bar \xi')\to
(\Sigma,\mathcal{F},\xi)$$

\begin{theorem}
Let $(\Sigma',\mathcal{L}',\xi')$ be an $ap$-subsystem of
$(\Sigma,\mathcal{L},\xi)$, generated by $a$. We now define the
following maps:
$$m:\Sigma' \to \Sigma:p \mapsto p$$
$$n:\mathcal{L}\to \mathcal{L}':c \mapsto a \wedge c$$
then $(m,n):(\Sigma',\mathcal{L}',\xi')\to
(\Sigma,\mathcal{L},\xi)$ is a morphism in the category of state
property systems which reduces to the canonical inclusion between
the underlying closure spaces.
\end{theorem}

\noindent
Proof:
We have to show that for $c \in {\cal L}$ and $p' \in \Sigma' : c
\in \xi(m(p')) \Leftrightarrow n(c) \in \xi'(p')$. Let's start
with $ c \in \xi(m(p')) \Leftrightarrow c\in \xi(p')
\Leftrightarrow c\in \xi'(p')$. Because $\kappa(a)=\Sigma'$ we
know that $a\in \xi'(p')=\xi(p')$, therefore $n(c)=c\wedge a\in
\xi'(p')$. Conversely, if $n(c)=c\wedge a\in \xi'(p')$ then $p'
\in \kappa'(c\wedge a)=\kappa'(c) \cap \kappa'(a)=\kappa'(c) \cap
\Sigma'=\kappa'(c)$ therefore $c \in \xi'(p)$.
\qed

\medskip
\noindent
We shall apply these results to the pure nonclassical state property systems
$(\Sigma_\omega,{\cal L}_\omega,\xi_\omega),\omega \in \Omega$ that we have
introduced in the
previous section. Recall that we started with a state property system $\sps$
with associated closure space $(\Sigma,\kappa(\mathcal{L}))$. By means of the
connection relation on $(\Sigma,\kappa(\mathcal{L}))$ we obtained a partition
$\Omega=\{\omega(p)=K_{\bf Cls}(p)|p\in \Sigma\}$ of $\Sigma$. Moreover each
$w\in \Omega$ with $\omega=\omega(p)=K_{\bf Cls}(p)$ was a closed subset of
$(\Sigma,\kappa(\mathcal{L}))$. Hence there was an $a=s(\omega)$ such that
$\kappa(a)=\omega$. We will now use this property $a=s(\omega)$ to create
an $ap$-subsystem.
$$\Sigma'=\kappa(a)=\omega$$
$$\mathcal{L}'=[0,a]=[0,s(\omega)]$$
$$\xi'=\xi_{|\Sigma'}^{|\mathcal{L}'}:p'\mapsto \xi(p)\cap \mathcal{L}'$$
We easily see that for an $\omega\in \Omega$ this $ap$-subsystem is in fact
$(\Sigma_\omega,{\cal L}_\omega,\xi_\omega)$. Let
$$m:\Sigma_\omega \to \Sigma:p \mapsto p$$
$$n:\mathcal{L}\to \mathcal{L}_\omega:c \mapsto s(\omega) \wedge c$$
then $(m,n):(\Sigma',\mathcal{L}',\xi')\to (\Sigma,\mathcal{L},\xi)$ is a
morphism in the category of state property systems which reduces to the
canonical inclusion between the underlying closure spaces. In this way
$(\Sigma_\omega,{\cal L}_\omega,\xi_\omega),\omega \in \Omega$ is always an
$ap$-subsystem of $\sps$.

\section{The D-classical Part of a State Property System}

In this section we want to show how it is possible to extract the
d-classical part of a state property system. First of all we have to
define the d-classical property lattice related to the entity $S$
that is described by the state property system $(\Sigma,{\cal
L},\xi)$.

\begin{definition} [D-classical Property Lattice]
Consider a state property system $(\Sigma,{\cal L},\xi)$. We call
${\cal C}' = \{\wedge_ia_i \vert a_i \ {\rm is \ a \ d-classical \
property }\}$ the d-classical property lattice corresponding to the
state property system $(\Sigma,{\cal L},\xi)$.
\end{definition}

\begin{theorem}${\cal C}'$ is a complete lattice with the partial order
relation
and infimum inherited from ${\cal L}$ and the supremum defined as
follows: for $a_i \in {\cal C}'$,  $\vee_ia_i = \wedge_{b \in
{\cal C}', a_i \le b\ \forall i}\ b$.
\end{theorem}

\noindent
Remark that the supremum in the lattice ${\cal C}'$ is not the one
inherited from ${\cal L}$.

\begin{theorem}
Consider a state property system $\sps$. Let $\xi'(q)=\xi(q)\cap
{\cal C}'$ for $q\in \Sigma$, then $(\Sigma,\mathcal{C}',\xi')$ is
a state property system which we shall refer to as the d-classical
part of $\sps$.
\end{theorem}

\noindent
Proof:
Clearly $0 \not\in \xi'(p)$ for $p\in \Sigma$. Consider $a_i \in
\xi'(p)\ \forall i$. Then $a_i \in \xi(p) \cap {\cal C}' \ \forall
i$, from which follows that $\wedge_ia_i \in \xi(p) \cap {\cal
C}'$ and hence $\wedge_ia_i \in \xi'(p)$. Consider $a, b \in {\cal
C}'$. Let us suppose that $a \leq b$ and consider $r \in \Sigma$
such that $a \in \xi'(r)$. This means that $a \in \xi(r) \cap
{\cal C}'$. From this follows that $b \in \xi(r) \cap {\cal C}'$
and hence $b \in \xi'(r)$. On the other hand let us suppose that
$\forall r \in \Sigma:a \in \xi'(r) {\rm \ then}\ b \in \xi'(r)$.
Since $a, b \in {\cal C}'$, this also means that $\forall r \in
\Sigma:a \in \xi(r) {\rm \ then}\ b \in \xi(r)$. From this follows
that $a \le b$.
\qed

\medskip
\noindent
Since $(\Sigma,\mathcal{C}',\xi')$ is a state property system, it
has a corresponding closure space $(\Sigma,\kappa(\mathcal{C}'))$.
In order to check some property of this space we introduce the
following concepts.

\begin{definition} [Weakly Zero-dimensional Closure Space]
Let $(X,\mathcal{F})$ be a closure space and $\mathcal{B}\subset
\mathcal{F}$. $\mathcal{B}$ is called a base of $(X,\mathcal{F})$
iff $\forall F\in \mathcal{F}:\exists B_i\in \mathcal{B}:F=\cap
B_i$. $(X,\mathcal{F})$ is called weakly zero-dimensional iff
there is a base consisting of clopen sets.
\end{definition}

\begin{theorem}
The closure space $(\Sigma,\kappa(\mathcal{C}'))$ corresponding to
the state property system $(\Sigma,\mathcal{C}',\xi')$  is weakly
zero-dimensional.
\end{theorem}

\noindent
Proof:
To see this recall that $a$ is classical iff $\kappa(a)$ is clopen
in $(\Sigma,\kappa(\mathcal{L}))$, hence $\kappa(\mathcal{C}')$ is
a family of closed sets on $\Sigma$ which consists of all
intersections of the clopen sets of
$(\Sigma,\kappa(\mathcal{L}))$.
\qed

\medskip
\noindent
In general $(\Sigma,\mathcal{C}',\xi')$ does not need to be
atomistic, hence it is different from the totally classical state
property system $(\Omega,\mathcal{C},\eta)$ associated with
$\sps$. To illustrate this we give an example.

\begin{figure*}[ht]
\centering
\includegraphics[width=4cm,height=4cm]{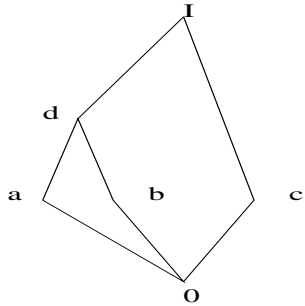}
\caption{The lattice $\mathcal{L}$}
\label{fig1}
\end{figure*}

\noindent
Let's consider the following state property system.
\begin{eqnarray*}
\Sigma &=& \{p,q,r,s,t\} \\
{\cal L} &=& \{0,a,b,c,d,I\} \\
\xi &:& \Sigma \to {\cal P}({\cal L})
\end{eqnarray*}
with $\xi(p) = \xi(q) = \{b,d,I\}, \xi(r) = \{a,d,I\}$ and
$\xi(s) = \xi(t) = \{c,I\}$. The structure for the lattice $\mathcal{L}$ is
given by figure \ref{fig1}.

\begin{figure*}[ht]
\centering
\includegraphics[width=5cm,height=5cm]{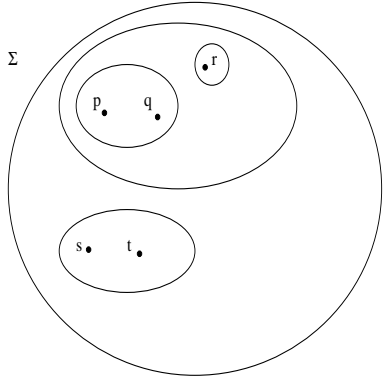}
\caption{The closure space $\Sigma,\kappa(\mathcal{L})$}
\label{fig2}
\end{figure*}

\noindent
The corresponding closure space (see figure \ref{fig2}) is
\begin{eqnarray*}
\Sigma &=& \{p,q,r,s,t\} \\
\kappa({\cal L}) &=& \{\emptyset, \{r\}, \{p,q\},\{s,t\},\{p,q,r\}, \Sigma\}
\end{eqnarray*}

\noindent
Determining the connectedness components in this closure space, we find the
following:
$$K_{\bf Cls}(p) = K_{\bf Cls}(q) = \{p,q\} $$
$$K_{\bf Cls}(r) = \{r\} $$
$$K_{\bf Cls}(s) = K_{\bf Cls}(t) = \{s,t\}$$
We have three pure nonclassical state property systems:
$(\Sigma_{{\omega}_1},{\cal L}_{{\omega}_1},\xi_{{\omega}_1})$,
$(\Sigma_{{\omega}_2},{\cal L}_{{\omega}_2},\xi_{{\omega}_2})$
and \\ $(\Sigma_{{\omega}_3},{\cal L}_{{\omega}_3},\xi_{{\omega}_3})$.
$$\begin{array}{ll}
\Sigma_{{\omega}_1} =\{p,q\},
&{\cal L}_{{\omega}_1} = [0,b] \\
\Sigma_{{\omega}_2} =\{r\},
&{\cal L}_{{\omega}_2} = [0,a] \\
\Sigma_{{\omega}_3} =\{s,t\},
&{\cal L}_{{\omega}_3} = [0,c]
\end{array}$$
$$\begin{array}{l}
\xi_{{\omega}_1}(p) =  \xi_{{\omega}_1}(q) = \{b\} \\
\xi_{{\omega}_2}(r) = \{a\} \\
\xi_{{\omega}_3}(s) =  \xi_{{\omega}_3}(t) = \{c\}
\end{array}$$
The atomistic totally classical state property system $(\Omega,{\cal
C},\eta)$ is given by:
\begin{eqnarray*}
{\Omega} &=& \{\{p,q\}, \{r\}, \{s,t\}\} \\
{\cal C} &=& {\cal L} \\
\eta &:& \Omega \to {\cal P}({\cal C})
\end{eqnarray*}
where $\eta(\{p,q\}) = \{b,d,1\}$, $\eta(\{r\}) = \{a,d,1\}$ and
$\eta(\{s,t\}) = \{c,1\}$.
The classical part is given by  $(\Sigma,{\cal C}',\xi')$ where
$$\xi'(p) = \xi(p) \cap {\cal C}'\hbox{ for } p \in \Sigma$$
$${\cal C}' = \{0,c,d,I\}$$

\section*{Acknowledgments}

Part of the research for this article took place in the framework of the
bilateral Flemish-Polish project 127/E-335/S/2000. D. Deses is Research Assistent at the
FWO Belgium.

\end{document}